# MICROCONTROLLER BASED AVR HAZARDOUS GAS DETECTION SYSTEM USING IOT

Ram Prasad

Department of Mathematics, VTU University, Bangalore, India

## ABSTRACT

*MQ-6 Semiconductor Sensor for Combustible Gas detection is a Sensitive Gas sensor. The sensitive material of this MQ-6 gas sensor is SnO2, which works with lower conductivity in clean air. When the target combustible gas exist, the sensors conductivity is higher along with the gas concentration rising. As the conductivity increases the current in the circuit of the sensor increases which results in lower sensor resistance. This change is used to correspond, the output signal of gas concentration. MQ-6 gas sensor has high sensitivity to Methane, Propane and Butane and could be used to detect both Methane and Propane. The sensor could be used to detect different combustible gas especially Methane, it is with low cost and suitable for different application.*

## KEYWORDS

*Wireless Network, MQ-6 gas sensor, SnO2, internet of thing,*

## 1. INTRODUCTION

A number of reviews on the subject of gas leakage detection techniques were done in the past either as part of research papers/technical reports on a certain leak detection method and other gas related subjects. In the year of 2008, LIU zhen-ya, WANG Zhen-dong and CHEN Rong, [1] "Intelligent Residential Security Alarm and Remote Control System Based On Single Chip Computer", the paper focuses on, Intelligent residential burglar alarm, emergency alarm, fire alarm, toxic gas leakage remote automatic sound alarm and remote control system, which is based on 89c51 single chip computer or 8051 microcontroller. The system that they design was used to send a message to emergency number provided and as well as call the police hotline number for emergency help.

In the year of 2006, Ioan Lita, Ion Bogdan Cioc and Daniel Alexandru Visan, [2] "A New Approach of Automatic Localization System Using GPS and GSM/GPRS Transmission", this paper focuses on, a low cost automotive localization system using GPS and GSM-SMS services, which provides the position of the vehicle on the driver's or owner's mobile phone as a short message (SMS) on his request. The system can be integrated with the car alarm system which alerts the owner, on his mobile phone, about the events that occurs with his car when it is parked. Or sends SMS to the relatives to provide fast emergency if any accident is happened.

In the year 2000, K. Galatsis, W. Woldarsla, Y.X. Li and K. Kalantar-zadeh, [3] "A Vehicle air quality monitor using gas sensors for improved safety", this paper focuses on A vehicle cabin air quality monitor using carbon monoxide (CO) and oxygen (02) gas sensors has been designed, developed and on-road tested. As of today the use of Air Conditioner (A/C) in the cars is more often this is dangerous to outer environment causing Global Warming like problems but as well as it affects the inner environment of a car. It causes problems like decrease in the oxygen level around 15% and increase in the level of harmful gases like Carbon mono oxide. The continuous monitoring of these gases increase vehicle safety and an alert can be made to let the passengers know that the concentration of the gases has reached their threshold value and it will be dangerous to further use the exhaust or AC in car. Later, in the 2002 they published another paper, [4] "Investigation of gas sensors for vehicle cabin air quality monitoring". In this they proposed the use of MOS (Metal oxide Semiconductor) Gas Sensor. Commercially available gas sensors are compared with fabricated Moo3 based sensors possessed comparable gas sensing properties.

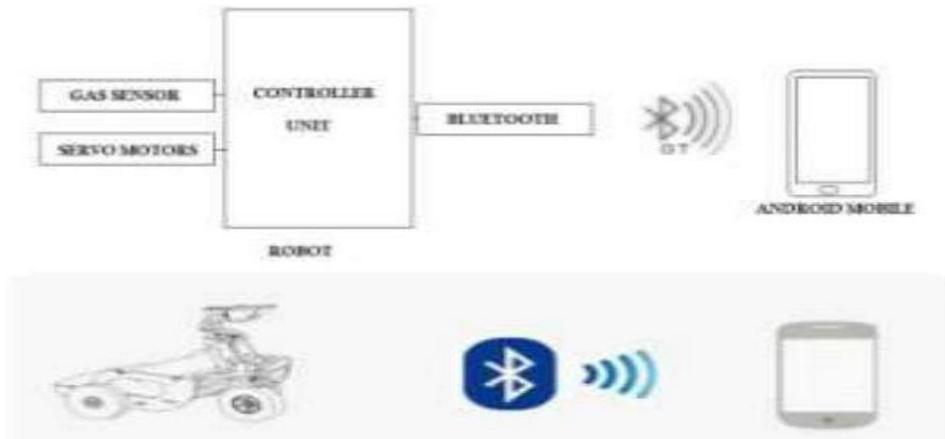

Figure 1. Bluetooth technology to control and monitor parameters driven by a robot

In year 2018; Zhao, W., Kamezaki, M., Yoshida, K., Konno, M., Onuki, A., & Sugano, S. [5] Worked on "An automatic tracked robot chain system for gas pipeline inspection and maintenance based on wireless relay communication". Mamatha, Sasritha, and Reddy, CS in the year 2017, deevelped an expert system and heuristics algorithm for cloud resource scheduling and introduced an android based automatic gas detection and indication robot[6, 7]. They proposed prototype depicts a mini mobile robot which is capable to detect gas leakage in hazardous places [8-12]. Whenever there is an event of gas leakage in a particular place the robot immediately read and sends the data to android mobile through wireless communication like Bluetooth. In this they develop an android application for android based smart phones which can receive data from robot directly through Bluetooth. The application warns with an indication whenever there is an occurrence of gas leakage and it can also be controlled by the robot movements via Bluetooth by using text commands as well as voice commands [13, 14].

In the paper [16], they introduced a robot and mobile application. Which made it is certain that an autonomous, mobile gas detection and leak localization robot is possible today and can significantly enhance safety. Shyamaladevi, [17] and her team; in their research article told about the project ARM7 based automated high performance system for LPG refill booking and leakage detection and methodology to make their project. In this, if there is a case of leakage, the resistance of the sensor decreases which increase its conductivity. The related output pulse is fed to microcontroller and which switches on the buzzer and exhaust fan to provide quick alert to house members [18, 19]. Microcontroller will send a message like "EMERGENCY ALERT: LPG gas leakage found in your home" to the required cell numbers via GSM module and the same will be displayed on LCD. The gas leakage detection system was proposed, designed and successfully implemented in this paper for home safety and industrial applications [20, 21]. Along with gas leakage detection, this system gives a fully automated approach towards the gas booking.

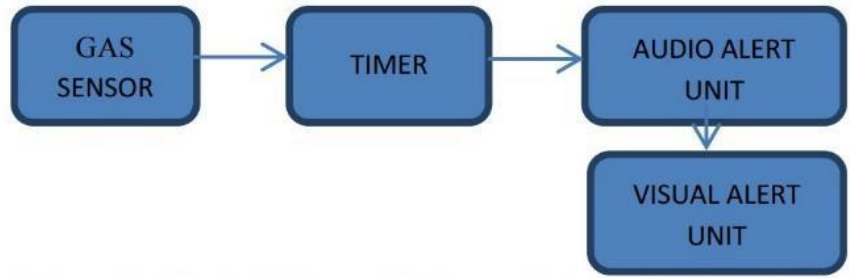

Figure 2. User Interactive Gas Leakage and Fire Alarm System

This project was implemented using the ARM 7 processor and simulated using the Keil software. The cost involved in developing the system is significantly low and is much less than the cost of gas detectors commercially available in the market. In year 2016, [22] "Dangerous gas detection using an integrated circuit and MQ-9" given by Falohun A.S., Oke A.O. and Abolaji B.M. This paper was focused on combustible gas detection. In this basically, they built an embedded design which includes typical input and output devices include switches, relays, solenoids, LEDs, small or custom LCD displays, radio frequency devices, and sensors for data such as temperature, humidity, light level etc [23-25]. Principle of operation proposed was the gas detector alarm system is designed with the intention to ensure that the event of gas is intelligently detected, promptly notified and interactively managed. It is built around a timer to accept input from the gas sensor, MQ-9, and activate a buzzer and set of led that alerts in the presence of gas. The sensor used is the MQ- 9 that specializes in gas detection equipment for carbon monoxide (CO) and Methane (CH4), LPG family and any other relevant industry or car assemblage.

According to the value received if that is above threshold, microcontroller will turn on LED and Buzzer and message is start viewing on the 16x2 LCD display [26-28]. Once few milliseconds delay, it conjointly sends the information over the internet for throwing gas out and continue to send messages as "Gas Leakage Detected" to the concerned mobile number. This information that is send over the server created on the internet and a Smartphone application can be used to notify [29]. The data on the server is displayed at a webpage for user.

The advantage of MQ-9 gas sensor is that it has; good sensitivity to CO/Combustible gas, high sensitivity to methane, propane and CO, long life and low cost and simple drive circuit. The enveloped MQ-9 has 6 pins, 4 of which are used to fetch signals, and other 2 for providing heating current. Once powered, the output of the sensor is normally HIGH but goes LOW when gas is sensed.

## 2. ARCHITECTURAL MODEL

Embedded Systems is defined in many ways. Few definitions such as, "An embedded system is a microprocessor based system that is built to control a function or a range of functions". An embedded system is some combination of computer hardware and software, either fixed in capability or programmable, that is designed for a specific function or for specific functions within a larger system. In the project, both the functions transmitting data over internet and sending text message to User mobile number, are done wirelessly using GSM module. MQ-6 Gas sensor is used to detect Hazardous Gases which are combustible in nature. The power supply of the project is regulated as 5V, supplied by a DC battery. The programming languages used for developing the software to the AVR Microcontroller is Embedded C as well as assembly language. The PROTEUS is used to stimulate the project on software.

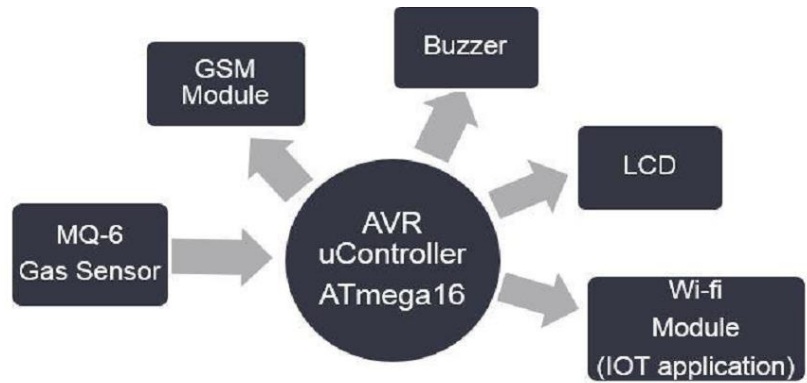

Figure 3. The Proposed Model Block Diagram

## 3. SYSTEM ASPECTS AND ACHIEVEMENTS

In this proposed model the authors want to achieve five aspects:

- **Design of Embedded System:** In this authors are using the AVR microcontroller that control all the module and other components and peripheral devices connected to it. Some of them work as input units and others as output units.
- **GSM Module**: GSM module is used to send the message of gas leakage to the emergency number provided in the program. This module is also used to setup an internet connection and use for IOT Applications. The data fetched from the Gas Sensor is uploaded to a virtual server which can be displayed on a webpage.
- **Alerting System**: This part includes a buzzer and a LCD Display. Buzzer is used to work as an alarm and LCD is being used to Display the Alert message and other details required to show.
- **Sensor Module**: This module is use to sense the gas leakage. In this proposed module they use a gas sensor MQ 6 to perform the leakage detection operation. This is the main component of the whole system. This is a Semiconductor type Gas sensor in which the conductivity increases as the gas concentration increases. To sense the concentration of gases $SnO_2$ and Au layer is used. The increases conductivity results in decrease in the Sensor load resistance, this change is measured by the controller. If the concentration goes beyond threshold it sends signal to controller which starts the tasks assigned by the programmer.
- **Software Used**: Two types of software that are used in developing this system are Atmel Studio and Proteus. Atmel Studio is used for programming. Embedded C is generally used to program microcontrollers. This studio generates a HEX code file which is required to burn in the microcontroller. On the other hand Proteus is simulation software. It is a general practice to run the program on Proteus before physically connecting the system to avoid failures.

## 4. SYSTEM IMPLEMENTATION

The hazardous gases like LPG and combustible gas were sensed by the MQ-6 gas sensor and are monitored by the AVR microcontroller and displayed in the LCD. In critical situation, that is when the LPG exceeds from normal level above 1000ppm and in the same way when the Propane exceeds the normal level of 10000ppm then an alarm is generated and a SMS is sent to the authorized user as an alerting system, which helps in faster diffusion of the critical situation. The prototype of the proposed system is shown in the below given figure.

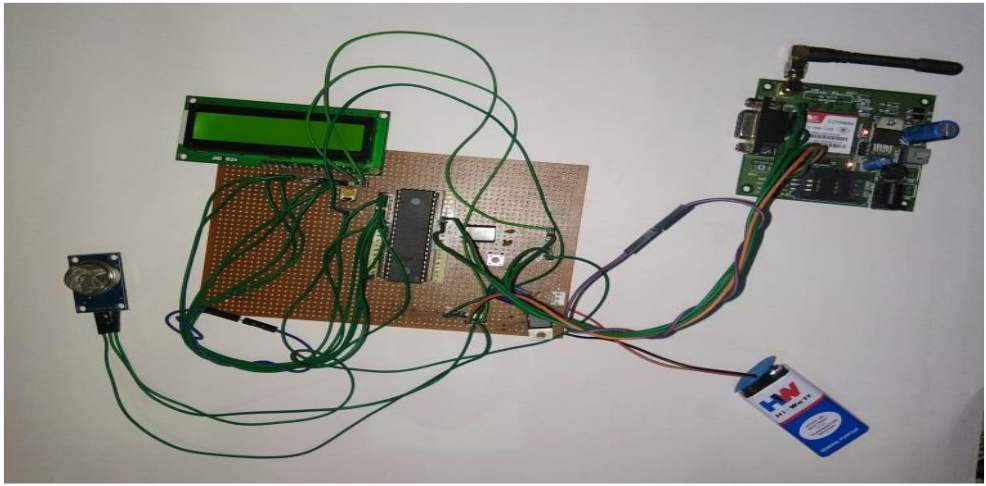

Figure 4. Gas detection unit result LCD display

## 5. FUTURE SCOPE

In the present days automation plays a vital role in the human life. Most of the day to day tasks are monitored and controlled by embedded controllers. Some of the applications that can be observed are:

**Mobile Application**: In this digital era when everyone has Smartphone. So people can easily connect to a mobile based application with the website to alert more efficiently in the users phone by giving the proper details. This Application will be directly connected to website for more instant results.

**Different Gases**: Environmental air quality is also becoming increasingly important and there will be many future requirements for low-cost, air quality monitoring sensors such as MEMS-based metal oxide Semiconductor sensors which are capable of monitoring pollutants such as ozone, carbon monoxide, nitrogen dioxide and ammonia.

**Infrared sensors**: The demands of fail-safe operation and higher reliability have caused many users to switch to infrared gas sensing. These types of sensor generally incorporate a pulsing source of infrared radiation that is absorbed by certain gases in proportion to the gas concentration. The infrared wavelength is chosen to suit a particular gas such as methane or carbon dioxide.

**Lower power and size**: The power consumed by pellistor and infrared types of gas sensors has limited their use in portable instrumentation to some extent due to battery capacities. Over the last few years rechargeable technology has developed greatly with battery chemistry migrating from Nickel Cadmium (Ni-Cd) to Nickel-Metal Hydride (NiMH) and now various forms of Lithium-Ion (Li-Ion), which offer the best power to weight ratio. The mobile phone and consumer electronic markets will continue to drive the development of new battery technologies with spin-off benefits to the portable gas sensor industry.

## 6. CONCLUSIONS

The importance of gas sensing is set to grow with increasing requirements for safety and environmental protection across many industries. The current range of gas sensing technologies has served us well but the future holds many new possibilities Power and size reductions and an improvement in ruggedness will allow a new generation of body worn devices A wide variety of leak detecting devices are available for gas pipelines. Some techniques have been improved since their first proposal and some new ones were designed as a result of advances in sensor manufacturing and computing power.

## REFERENCES

[1] Zhen-ya, L. I. U., Zhen-dong, W., Rong, C. H. E. N., & Xiao-feng, W. (2008). Intelligent residential security alarm and remote control system based on single chip computer. In 2008 3rd IEEE Conference on Industrial Electronics and Applications.

[2] Lita, I., Cioc, I. B., & Visan, D. A. (2006, May). A new approach of automobile localization system using GPS and GSM/GPRS transmission. In 2006 29th International Spring Seminar on Electronics Technology (pp. 115-119). IEEE.

[3] Galatsis, K., Wlodarski, W., Li, Y. X., & Kalantar-Zadeh, K. (2000). Vehicle cabin air quality monitor using gas sensors for improved safety. In COMMAD 2000 Proceedings. Conference on Optoelectronic and Microelectronic Materials and Devices (pp. 65-68). IEEE.

[4] Mamatha, E., Reddy, C. S., & Prasad, K. R. (2016). Antialiased Digital Pixel Plotting for Raster Scan Lines Using Area Evaluation. In Emerging Research in Computing, Information, Communication and Applications: ERCICA 2015, Volume 3 (pp. 461-468). Springer Singapore.

[5] Zhao, W., Kamezaki, M., Yoshida, K., Konno, M., Onuki, A., & Sugano, S. (2018, October). An automatic tracked robot chain system for gas pipeline inspection and maintenance based on wireless relay communication. In 2018 IEEE/RSJ International Conference on Intelligent Robots and Systems (IROS) (pp. 3978-3983). IEEE.

[6] Mamatha, E., Sasritha, S., & Reddy, C. S. (2017). Expert system and heuristics algorithm for cloud resource scheduling. Romanian Statistical Review, 65(1), 3-18.

[7] Mamatha, E., Saritha, S., Reddy, C. S., & Rajadurai, P. (2020). Mathematical modelling and performance analysis of single server queuing system-eigenspectrum. International Journal of Mathematics in Operational Research, 16(4), 455-468.

[8] Anand, K., Mamatha, E., Reddy, C. S., & Prabha, M. (2019). Design of neural network based expert system for automated lime kiln system. Journal Européen des Systèmes Automatisés, 52(4), 369-376.

[9] Mamatha, E., Reddy, C. S., & Anand, S. K. (2016). Focal point computation and homogeneous geometrical transformation for linear curves. Perspectives in Science, 8, 19-21.

[10] Mamatha, E., Saritha, S., & Reddy, C. S. (2016). Stochastic scheduling algorithm for distributed cloud networks using heuristic approach. International Journal of Advanced Networking and Applications, 8(1), 3009.

[11] Reddy, C. S., Janani, B., Narayanan, S., & Mamatha, E. (2016). Obtaining Description for Simple Images using Surface Realization Techniques and Natural Language Processing. Indian Journal of Science and Technology, 9(22), 1-7.

[12] Kumar, M. S., Mamatha, E., Reddy, C. S., Mukesh, V., & Reddy, R. D. (2017, September). Data hiding with dual based reversible image using sudoku technique. In 2017 International Conference on Advances in Computing, Communications and Informatics (ICACCI) (pp. 2166-2172). IEEE.

[13] Yadav, S. A., Sharma, S., Das, L., Gupta, S., & Vashisht, S. (2021, February). An Effective IoT Empowered Real-time Gas Detection System for Wireless Sensor Networks. In 2021 International Conference on Innovative Practices in Technology and Management (ICIPTM) (pp. 44-49). IEEE.

[14] Elliriki, M., Reddy, C. S., Anand, K., & Saritha, S. (2022). Multi server queuing system with crashes and alternative repair strategies. Communications in Statistics-Theory and Methods, 51(23), 8173-8185.

[15] Sharma, P., & Kharb, L. (2022). IoT-Enabled Hazardous Gas Leakage Detection System for Citizen's Safety. In Internet of Things (pp. 257-270). CRC Press.

[16] Elliriki, M., Reddy, C. C. S., & Anand, K. (2019). An efficient line clipping algorithm in 2D space. Int. Arab J. Inf. Technol., 16(5), 798-807.
[17] Shyamaladevi, S., Rajaramya, V. G., Rajasekar, P., & Ashok, P. S. (2014). ARM 7 based automated high performance system for LPG refill booking & Leakage detection. International journal of engineering research, science and technology (IJERST), 3(2).
[18] Yang, Z., Liu, M., Shao, M., & Ji, Y. (2011). Research on leakage detection and analysis of leakage point in the gas pipeline system. Open journal of safety science and technology, 1(03), 94-100.
[19] Saritha, S., Mamatha, E., Reddy, C. S., & Rajadurai, P. (2022). A model for overflow queuing network with two-station heterogeneous system. International Journal of Process Management and Benchmarking, 12(2), 147-158.
[20] Mamatha, E., Reddy, C. S., & Prasad, R. (2012). Mathematical modeling of markovian queuing network with repairs, breakdown and fixed buffer. i-Manager's Journal on Software Engineering, 6(3), 21.
[21] Shrivastava, A., Prabhaker, R., Kumar, R., & Verma, R. (2013). GSM based gas leakage detection system. International Journal of Technical Research and Applications, 1(2), 42-45.
[22] Falohun, A. S., Oke, A. O., Abolaji, B. M., & Oladejo, O. E. (2016). Dangerous gas detection using an integrated circuit and MQ-9. International Journal of Computer Applications, 135(7), 30-34.
[23] Saritha S, Mamatha E, Reddy CS, Anand K. A model for compound poisson process queuing system with batch arrivals and services. Journal Europeen des Systemes Automatises. 2019;53(1):81-6.
[24] Saritha, S., Mamatha, E., & Reddy, C. S. (2019). Performance measures of online warehouse service system with replenishment policy. Journal Europeen Des Systemes Automatises, 52(6), 631-638.
[25] Mamatha, E., Reddy, C. S., & Sharma, R. (2018). Effects of viscosity variation and thermal effects in squeeze films. In Annales de Chimie. Science des Materiaux (Vol. 42, No. 1, p. 57). International Information and Engineering Technology Association (IIETA).
[26] Bhattacharyya, A., Sharma, R., Hussain, S. M., Chamkha, A. J., & Mamatha, E. (2022). A numerical and statistical approach to capture the flow characteristics of Maxwell hybrid nanofluid containing copper and graphene nanoparticles. Chinese Journal of Physics, 77, 1278-1290.
[27] Sreelatha, V., Mamatha, E., Reddy, C. S., & Rajdurai, P. S. (2022). Spectrum Relay Performance of Cognitive Radio between Users with Random Arrivals and Departures. In Mobile Radio Communications and 5G Networks: Proceedings of Second MRCN 2021 (pp. 533-542). Singapore: Springer Nature Singapore.
[28] Saritha, S., Mamatha, E., Reddy, C. S., & Anand, K. (2019). A model for compound poisson process queuing system with batch arrivals and services. Journal Europeen des Systemes Automatises, 53(1), 81-86.
[29] Reddy, C. S., & Anand, K. (2022). Sensor Signal Processing using High-Level Synthesis and Internet of Things with a Layered Architecture, International Journal on AdHoc Networking Systems (IJANS) Vol. 12, No. 4,